# Influence of bounce resonance effects on the cyclotron wave instabilities in dipole magnetospheric plasmas with anisotropic temperature

## N.I. Grishanov[1], M.A. Raupp[1], A.F.D. Loula[1], J. Pereira Neto[2]

[1]*Laboratório Nacional de Computação Científica, Petrópolis, RJ, BRASIL*
[2]*Universidade do Estado do Rio de Janeiro, Rio de Janeiro, RJ, BRASIL*

**Abstract:** Dispersion equations have been derived for field-aligned cyclotron waves in dipole magnetospheric plasmas with anisotropic temperature. Vlasov equation is solved for trapped particles with both the bi-Maxwellian and bi-Lorentzian distribution functions accounting for the cyclotron and bounce resonances. New transverse permittivity elements are expressed by the summation of bounce-resonant terms including the double integration in velocity space, the resonant denominators, and the phase coefficients. To have some analogy with the linear theory of cyclotron waves in the straight magnetic field, we assume that the *n*-th harmonic of the electric field gives the main contribution to the *n*-th harmonic of the current density, and the connection of the left- and right-hand waves is small. In this case, the dispersion equations for cyclotron waves have the simplest form and are suitable to annualise the instabilities of both the electron- and ion-cyclotron waves accounting for the bounce resonance effects.

## 1. Introduction and plasma model

Cyclotron waves are an important component of the magnetospheric environment. As is well known, the energetic particles (electrons, protons, heavy ions) with anisotropic temperature (pressure) can excite a wide class of cyclotron wave instabilities. Kinetic theory of electromagnetic cyclotron waves/instabilities in the straight magnetic field case is developed quite fully, see e.g. Refs. [1-3] and bibliography therein. However, for magnetospheric plasmas, these instabilities should be analysed by solving the Vlasov-Maxwell's equations taking into account a two-dimensional nonuniformity of the geomagnetic field and bounce-resonant wave-particle interactions [4-7].

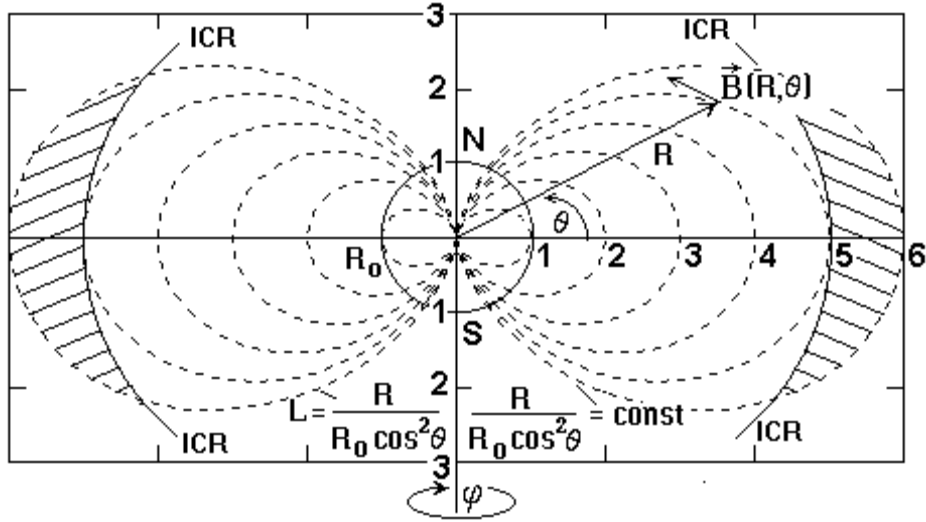

*Fig. 1. The point-dipole magnetic field configuration for the Earth's magnetosphere; ICR-ICR lines are the lines of the constant geomagnetic field.*

In this paper, we derive the dispersion relation for the field-aligned cyclotron waves in an axisymmetric dipole magnetosphere, where the module of an equilibrium magnetic field, in the spherical coordinates ($R, \theta, \varphi$), is (in the point dipole approximation)

$$B(R,\theta) = B_0 \left( \frac{R_0}{R} \right)^3 \sqrt{1 + 3\sin^2\theta} \quad . \tag{1}$$

Here, $R_0$ is the radius of the Earth, $R$ is the geocentric distance, $\theta$ is the geomagnetic latitude, $B_0$ is the Earth's equatorial magnetic field, see Fig. 1. There are considered the collisionless plasmas including the particles with anisotropic temperature, where charged particles are magnetically trapped having the two steady-state distribution functions in velocity space: 1) bi-Maxwellian, and/or 2) bi-Lorentzian.

## 2. Reduced Vlasov equation

To evaluate the transverse dielectric permittivity elements we solve the Vlasov equation for trapped particles with anisotropic temperature using a standard method of switching to new variables associated with the conservation integrals of the energy: $v_\parallel^2 + v_\perp^2 = const$, the magnetic moment: $v_\perp^2 / 2B = const$, and the **B**-field line equation: $R / \cos^2\theta = const$.

Introducing the variables

$$v = \sqrt{v_\parallel^2 + v_\perp^2}\,, \qquad \mu = \frac{v_\perp^2 \cos^6\theta}{v^2 \sqrt{1+3\sin^2\theta}}\,, \qquad L = \frac{R}{R_0 \cos^2\theta} \tag{2}$$

(instead of $v_\parallel, v_\perp, R$), the perturbed distribution function can be found as

$$f(t,R,\theta,\varphi,v_\parallel,v_\perp,\alpha) = \sum_{s}^{\pm 1}\sum_{l}^{\pm\infty} f_l^s(\theta,L,v,\mu)\exp(-i\omega t + im\varphi + il\alpha) \tag{3}$$

where $\alpha$ is the gyrophase angle in velocity space. The linearized Vlasov equation for (interesting us) harmonics $f_{\pm 1}^s$ can be rewritten in the form

$$\frac{\sqrt{1-\mu b(\theta)}}{\cos\theta\sqrt{1+3\sin^2\theta}}\frac{\partial f_l^s}{\partial\theta} - is\frac{R_0 L}{v}\left[\omega + l\frac{\omega_{co}}{L^3}b(\theta)\right]f_l^s = Q_l^s\,, \qquad l = \pm 1 \tag{4}$$

where

$$Q_l^s = \frac{eR_0 L}{Mv_{T\parallel}^2}\sqrt{\mu}F_0\left[s\frac{E_l}{\sqrt{b(\theta)}}\left[b(\theta)-1+\frac{T_\parallel}{T_\perp}\right] - \frac{iv\sqrt{1-\mu b(\theta)}(1-T_\parallel/T_\perp)}{\omega R_0 L\cos\theta\sqrt{1+3\sin^2\theta}}\frac{\partial}{\partial\theta}\frac{E_l}{\sqrt{b(\theta)}}\right] \tag{5}$$

$$F_0 = \frac{N(L)}{\pi^{1.5}v_{T\parallel}v_{T\perp}^2}\exp\left\{-\frac{v^2}{v_\parallel^2}\left[1-\mu\left(1-\frac{T_\parallel}{T_\perp}\right)\right]\right\}\,, \qquad b(\theta) = \frac{\sqrt{1+3\sin^2\theta}}{\cos^6\theta} \tag{6}$$

$$E_l = E_n - ilE_b\,, \qquad v_{T\parallel}^2 = \frac{2T_\parallel}{M}\,, \qquad v_{T\perp}^2 = \frac{2T_\perp}{M}. \tag{7}$$

Here, $E_n$ and $E_b$ are, respectively, the normal and binormal perturbed electric field components relative to **B**; $F_0$ is the bi-Maxwellian distribution function of plasma particles with density $N$, parallel and perpendicular temperature $T_\parallel$ and $T_\perp$, charge $e$ and mass $M$. By the indexes $s = \pm 1$ we distinguish the particles with positive and negative values of the parallel velocity

$$v_\parallel = sv\sqrt{1-\mu b(\theta)}\,, \tag{8}$$

relative to **B**. In Eq. (4) we have neglected the drift corrections assuming that the wave frequency $\omega$ is much larger than the drift frequency, that is valid when $mv_{T\perp}^2 L^2 / (v_{T\parallel}R_0\omega_{co}) \ll 1$, where $\omega_{co} = eB_0 / Mc$, and $m$ is the azimuthal wave number over the $\varphi$ (east-west) direction. Moreover, deriving the dispersion equation, as well as Eq. (4), we assume that the plasma is perfectly conducting, i.e., $E_\parallel = 0$, and the ordinary ($E_{-1} = E_n + iE_b$) and extraordinary ($E_{+1} = E_n - iE_b$) waves are connected weakly.

Depending on $\mu$ and $\theta$, the domain of the perturbed distribution functions is defined by the inequalities

$$\mu_0 \le \mu \le 1 \quad \text{and} \quad -\theta_t(\mu) \le \theta \le \theta_t(\mu), \tag{9}$$

where

$$\mu_0 = \frac{1}{b(\arccos\dfrac{1}{\sqrt{L}})} = \frac{1}{L^{2.5}\sqrt{4L-3}}, \tag{10}$$

and $\pm\theta_t(\mu)$ are the local mirror (or turning, or reflection) points for trapped particles at a given (by $L$) magnetic field line, which are defined by the condition $v_\parallel(\pm\theta_t) = 0$. Any untrapped particle with $\mu \le \mu_0$ cannot survive more than one half of the bounce-time and will be precipitated into the atmosphere/ionosphere.

Since the trapped particles, with a given (by $\mu$) pitch angle, execute the bounce-periodic motion, the solution of Eq. (4) can be found as

$$f_l^s(\theta, L, v, \mu) = \sum_{p=-\infty}^{+\infty} f_l^s(L, v, \mu)\exp\left[ip\frac{2\pi}{\tau_b}\tau(\theta) + isl\frac{R_0\omega_{co}}{L^2 v}C(\theta)\right] \tag{11}$$

where

$$\tau(\theta) = \int_0^\theta \frac{\cos\eta\sqrt{1+3\sin^2\eta}}{\sqrt{1-\mu b(\eta)}}d\eta, \qquad \tau_b = \tau_b(\mu) = 4\tau(\theta_t) \tag{12}$$

$$C(\theta) = \int_0^\theta b(\eta)\frac{\cos\eta\sqrt{1+3\sin^2\eta}}{\sqrt{1-\mu b(\eta)}}d\eta - \bar{b}\,\tau(\theta), \qquad \bar{b} = \frac{4}{\tau_b}\int_0^{\theta_t} b(\theta)\frac{\cos\theta\sqrt{1+3\sin^2\theta}}{\sqrt{1-\mu b(\theta)}}d\theta. \tag{13}$$

The perturbed distribution functions, accounting for Eq. (11), satisfy automatically the corresponding boundary conditions for the trapped particles, namely, the continuity of the distribution functions at the reflection points $\pm\theta_t$: $f_l^{+1}(\pm\theta_t) = f_l^{-1}(\pm\theta_t)$, or the same $f_l^s(\tau(\theta)) = f_l^s(\tau(\theta) + \tau_b)$. In our notations, the bounce frequency of the trapped particles with temperature $T$ and given parameter $\mu$ is defined as $\omega_b = 2\pi v_{T\parallel}/(R_0 L\,\tau_b)$.

After solving Eq. (4), the contribution of trapped particles to the (two-dimensional) transverse current density component, $j_{\pm 1}(\theta, L)$, can be expressed as

$$j_l(\theta, L) = \frac{\pi e}{2}b^{1.5}(\theta)\sum_{s}^{\pm 1}\int_0^\infty v^3 dv\int_{\mu_0}^{1/b(\theta)}\frac{\sqrt{\mu}f_l^s(\theta, L, v, \mu)}{\sqrt{1-\mu b(\theta)}}d\mu, \qquad l = \pm 1 \tag{14}$$

Note that the normal and binormal (to $\mathbf{B}$) current density components in our notation are equal to $j_n = j_{+1} + j_{-1}$ and $j_b = i(j_{+1} - j_{-1})$, respectively.

## 3. Transverse permittivity elements

To solve the wave (or Maxwell's) equations let us expand preliminary the perturbed values in a Fourier series over $\theta$. In particular, for the transverse components of the current density, $j_l$, and electric field, $E_l$, we have:

$$\frac{j_l(\theta, L)}{\sqrt{b(\theta)}} = \sum_{n}^{\pm\infty} j_l^{(n)}(L)\exp\left[i\pi n\frac{\lambda(\theta)}{\lambda_o}\right], \qquad \frac{E_l(\theta, L)}{\sqrt{b(\theta)}} = \sum_{n'}^{\pm\infty} E_l^{(n')}(L)\exp\left[i\pi n'\frac{\lambda(\theta)}{\lambda_o}\right] \tag{15}$$

where

$$\lambda(\theta) = \frac{R_0 L}{2\sqrt{3}} \left[ \sqrt{3}\sin\theta\sqrt{1+3\sin^2\theta} + \ln\left(\sqrt{3}\sin\theta + \sqrt{1+3\sin^2\theta}\right) \right] \tag{16}$$

and $\lambda_o = \lambda(\arccos\frac{1}{\sqrt{L}})$ is the half-length of a given by L magnetic field line. This procedure converts the operator, representing the dielectric tensor, into a matrix whose elements are calculated independently of the solutions of Maxwell's equations. As a result, there is following connection for harmonics $j_l^{(n)}$ and $E_l^{(n')}$:

$$\frac{4\pi i}{\omega} j_l^{(n)}(L) = \sum_{n'}^{\pm\infty} \varepsilon_l^{n,n'}(L) \cdot E_l^{(n')}(L) \tag{17}$$

and the contribution of a given kind of plasma particles to the transverse permittivity elements, $\varepsilon_l^{n,n'}(L)$, after the *s*-summation, is

$$\varepsilon_l^{n,n'} = \frac{\omega_{po}^2 L^2 R_0^2 T_\parallel}{8\omega\pi^{1.5}\lambda_o v_{T\parallel} T_\perp} \sum_{p=-\infty}^{\infty} \int_{\mu_0}^{1} \mu \ d\mu \int_{-\infty}^{\infty} u^4 \frac{D_{p,l}^n(u,\mu) G_{p,l}^{n'}(u,\mu)}{pu - Z_l(\mu)} \exp\left[-u^2\left(1-\mu\left(1-\frac{T_\parallel}{T_\perp}\right)\right)\right] du \tag{18}$$

where

$$\omega_{po}^2 = \frac{4\pi Ne^2}{M}, \qquad u = \frac{v}{v_{T\parallel}}, \qquad Z_l(\mu) = \frac{1}{\omega_b}\left(\omega + l\frac{\omega_{co}}{L^3}\bar{b}\right), \tag{19}$$

$$G_{p,l}^n(u,\mu) = \int_{-\theta_t}^{\theta_t} \left(b(\theta) - 1 + \frac{T_\parallel}{T_\perp} + \frac{\pi nuv_{T\parallel}}{\omega\lambda_o}\left(1-\frac{T_\parallel}{T_\perp}\right)\sqrt{1-\mu b(\theta)}\right) \times$$

$$\times \cos\left[\frac{n\pi}{\lambda_o}\lambda(\theta) - p\frac{2\pi}{\tau_b}\tau(\theta) - \frac{lR_0\omega_{co}}{L^2 uv_{T\parallel}}C(\theta)\right]\frac{\cos\theta\sqrt{1+3\sin^2\theta}}{\sqrt{1-\mu b(\theta)}} d\theta +$$

$$+ (-1)^p \int_{-\theta_t}^{\theta_t} \left(b(\theta) - 1 + \frac{T_\parallel}{T_\perp} + \frac{\pi nuv_{T\parallel}}{\omega\lambda_o}\left(1-\frac{T_\parallel}{T_\perp}\right)\sqrt{1-\mu b(\theta)}\right) \times$$

$$\times \cos\left[\frac{n\pi}{\lambda_o}\lambda(\theta) + p\frac{2\pi}{\tau_b}\tau(\theta) + \frac{lR_0\omega_{co}}{L^2 uv_{T\parallel}}C(\theta)\right]\frac{\cos\theta\sqrt{1+3\sin^2\theta}}{\sqrt{1-\mu b(\theta)}} d\theta, \tag{20}$$

$$D_{p,l}^n(u,\mu) = \int_{-\theta_t}^{\theta_t} \cos\left[\frac{n\pi}{\lambda_o}\lambda(\theta) - p\frac{2\pi}{\tau_b}\tau(\theta) - \frac{lR_0\omega_{co}}{L^2 uv_{T\parallel}}C(\theta)\right]b(\theta)\frac{\cos\theta\sqrt{1+3\sin^2\theta}}{\sqrt{1-\mu b(\theta)}} d\theta. \tag{21}$$

Thus, the transverse permittivity elements for electromagnetic waves (at the fundamental cyclotron frequency) in an axisymmetric dipole magnetospheric plasma are expressed by the *p*-summation of the bounce-resonant terms including the double integration in velocity space, the resonant denominators $pu-Z_l(\mu)$, and the phase coefficients $G_{p,l}^n(u,\mu)$ and $D_{p,l}^n(u,\mu)$. As follows from Eq. (17), due to a two-dimensional geomagnetic field nonuniformity, the whole spectrum of the electric field (by $\Sigma_{n'}^{\pm\infty}$) is present in the given (by *n*) current density harmonic. It should be noted that the bounce-resonance conditions, $pu-Z_l(\mu)=0$ for trapped particles in magnetospheric plasmas are entirely different from the corresponding expressions in the straight magnetic field case. Of course, as in the straight magnetic field, *l=1* corresponds to the effective resonant interaction of electrons with the extraordinary (or right-hand polarised) waves at the fundamental electron-cyclotron frequency, and *l=-1* corresponds to the resonant interaction of ions with an ordinary (or left-hand polarised) wave at the fundamental ion-cyclotron frequency. Note that there is no possibility to carry out the Landau integration over the particle energy $u = v/v_{T\parallel}$ (by introducing the plasma dispersion function) because the phase coefficients $G_{p,l}^{n'}(u,\mu)$ and $D_{p,l}^{n}(u,\mu)$ depend on $u$. As for the particles with isotropic temperature, i.e. if $T_{\parallel} = T_{\perp}$, the phase coefficients $G_{p,l}^{n}(u,\mu)$ can be reduced to $G_{p,l}^{n}(u,\mu) = D_{p,l}^{n}(u,\mu) + (-1)^{p} D_{-p,l}^{n}(-u,\mu)$.

As was noted above, Eq. (18) describes the contribution of any kind of the trapped particles to the transverse permittivity elements. The corresponding expressions for plasma electrons and ions can be obtained from (18) by replacing $T_{\parallel}$, $T_{\perp}$ (temperature), $N$ (density), $M$ (mass), $e$ (charge) by the electron $T_{\parallel e}$, $T_{\perp e}$, $N_e$, $m_e$, $e_e$ and ion $T_{\parallel i}$, $T_{\perp i}$, $N_i$, $M_i$, $e_i$ parameters, respectively.

## 4. Dispersion equation of the cyclotron waves

Since the cyclotron wave instabilities are an important contributor to cool plasma heating, it is possible (and interesting) to develop a two-dimensional numerical code to describe these processes in the Earth's magnetosphere with new dielectric tensor components, accounting for the bounce-resonant effects.

To have some analogy with linear theory[1-3] of cyclotron wave instabilities in the straight magnetic field, let us assume that the $n$-th harmonic of the electric field gives the main contribution to the $n$-th harmonic of the current density (*one-mode approximation*). In this case, for the field-aligned electromagnetic cyclotron waves (when $m=0, \partial/\partial L = 0, E_{\parallel} = 0, H_{\parallel} = 0$), there is following dispersion relation:

$$\left(\frac{n\pi c}{\lambda_o \omega}\right)^2 = 1 + 2\sum_{\sigma} \varepsilon_{l,\sigma}^{n,n}(L),\tag{22}$$

where $\sigma$ denotes the particle species (electron, proton, etc.). This equation is suitable to analyze the instability of the electron-cyclotron waves if $l=1$, and ion-cyclotron waves if $l=-1$. Note that, in our notation, the parallel wave vector is defined as $k_{\parallel} = n\pi/\lambda_o$, so that $n\pi c/(\lambda_o \omega)$ is the nondimensional parallel refractive index. Further, Eq. (22) should be resolved numerically for the real and imaginary parts of the wave frequency, $\omega = \mathrm{Re}\,\omega + \iota\,\mathrm{Im}\,\omega$, to define the conditions of the ion/electron cyclotron instabilities in the dipole magnetospheric plasmas with anisotropic temperature. As usual, the growth (damping) rate of the electromagnetic cyclotron waves, $\gamma = \mathrm{Im}\,\omega/\mathrm{Re}\,\omega$, is defined by the contribution of the resonant particles to the imaginary part of the transverse permittivity elements. The simplest estimation can be done assuming that $\mathrm{Im}\,\omega \ll \mathrm{Re}\,\omega$; in the case,

$$\gamma = \frac{\mathrm{Im}\,\omega}{\mathrm{Re}\,\omega} \approx -\left(\frac{\lambda_o\,\mathrm{Re}\,\omega}{\pi\,n\,c}\right)^2 \sum_{\sigma} \mathrm{Im}\,\varepsilon_{l,\sigma}^{n,n}(L).\tag{23}$$

To find the instability conditions (e.g., the thresholds) it is necessary to compare the wave growth rate (by the energetic particles with anisotropic temperature) and the corresponding damping rate (by the cold particles with isotropic temperature). The contributions of the resonant $\sigma$-particles to $\mathrm{Im}\,\varepsilon_{l,\sigma}^{n,n}$ can be readily derived from Eqs. (18) by using the well known residue (or Landau rule) method.

## 5. Lorentzian (or *kappa*) distribution function

Of course, an approach developed in sections 2-4 for magnetospheric plasmas with bi-Maxwellian distribution functions in velocity space can be applied as well for plasmas with the more general distributions including, e.g., the bi-Lorentzian distribution functions.

According to Ref. [3], the generalised steady-state bi-Lorentzian (or *kappa*) distribution functions can be expressed as

$$F_0 = \frac{N(L)}{\pi^{1.5}\vartheta_\parallel \vartheta_\perp^2 \kappa^{1.5}} \frac{\Gamma(1+\kappa)}{\Gamma(\kappa-0.5)}\left\{1+\frac{v^2}{\kappa\vartheta_\parallel^2}\left[1-\mu\left(1-\frac{T_\parallel}{T_\perp}\right)\right]\right\}^{-(1+\kappa)} \tag{24}$$

with the associated effective thermal speeds corresponding to $T_\parallel$ and $T_\perp$,

$$\vartheta_\parallel^2 = \frac{2\kappa-3}{\kappa}\frac{T_\parallel}{M}, \qquad \vartheta_\perp^2 = \frac{2\kappa-3}{\kappa}\frac{T_\perp}{M}, \tag{25}$$

where the parameter $\kappa$ is the spectral index (here takes positive values $\kappa \geq 2$); $\Gamma(x) = \int_0^\infty t^{x-1}\exp(-t)dt$ is the gamma function. It should be noted that the parameter $\kappa$ is a measure of the proportion of the high energy particles present in the distribution; typically for space plasmas, is found to be in the range $2 \leq \kappa \leq 6$. Moreover, the generalised bi-Lorentzian distribution contains the standard bi-Maxwellian distribution, Eq. (6), as a special case letting $\kappa \to \infty$.

To estimate the main contribution of the trapped particles with bi-Lorentzian distribution functions to the transverse permittivity elements we should solve again Eq. (4) where the right hand side is

$$\hat{Q}_l^s = \frac{eR_0 L\sqrt{\mu}(1+\kappa)F_0}{M\kappa\vartheta_\parallel^2\left[1+\dfrac{v^2}{\kappa\vartheta_\parallel^2}\left[1-\mu\left(1-\dfrac{T_\parallel}{T_\perp}\right)\right]\right]}\times$$

$$\times\left[\frac{sE_l}{\sqrt{b(\theta)}}\left[b(\theta)-1+\frac{T_\parallel}{T_\perp}\right]-\frac{iv\sqrt{1-\mu b(\theta)}(1-T_\parallel/T_\perp)}{\omega R_0 L\cos\theta\sqrt{1+3\sin^2\theta}}\frac{\partial}{\partial\theta}\frac{E_l}{\sqrt{b(\theta)}}\right]. \tag{26}$$

As a result, instead of Eq. (18), we have

$$\varepsilon_l^{n,n'} = \frac{\omega_{po}^2 L\, R_0\, T_\parallel\,(\kappa+1)\Gamma(\kappa+1)}{8\omega\pi^{1.5}\lambda_o T_\perp\sqrt{\kappa}\,\vartheta_\parallel\Gamma(\kappa-0.5)}\sum_{p=-\infty}^{\infty}\int_{\mu_0}^1 \mu\ d\mu\int_{-\infty}^{\infty}\frac{u^4\ \hat{D}_{p,l}^n(u,\mu)\ \hat{G}_{p,l}^{n'}(u,\mu)\ du}{[pu-\hat{Z}_l(\mu)]\left[1+u^2\left(1-\mu\left(1-\dfrac{T_\parallel}{T_\perp}\right)\right)\right]^{\kappa+2}} \tag{27}$$

where

$$\hat{G}_{p,l}^n(u,\mu) = \int_{-\theta_l}^{\theta_l}\left(b(\theta)-1+\frac{T_\parallel}{T_\perp}+\frac{\pi n u\sqrt{\kappa}\,\vartheta_\parallel}{\omega\ \lambda_o}\left(1-\frac{T_\parallel}{T_\perp}\right)\sqrt{1-\mu b(\theta)}\right)\times$$

$$\times\cos\left[\frac{n\pi}{\lambda_o}\lambda(\theta)-p\frac{2\pi}{\tau_b}\tau(\theta)-\frac{lR_0\omega_{co}}{L^2 u\sqrt{\kappa}\vartheta_\parallel}C(\theta)\right]\frac{\cos\theta\sqrt{1+3\sin^2\theta}}{\sqrt{1-\mu b(\theta)}}\,d\theta\ +$$

$$+(-1)^p\int_{-\theta_l}^{\theta_l}\left(b(\theta)-1+\frac{T_\parallel}{T_\perp}+\frac{\pi n u\sqrt{\kappa}\,\vartheta_\parallel}{\omega\ \lambda_o}\left(1-\frac{T_\parallel}{T_\perp}\right)\sqrt{1-\mu b(\theta)}\right)\times$$

$$\times\cos\left[\frac{n\pi}{\lambda_o}\lambda(\theta)+p\frac{2\pi}{\tau_b}\tau(\theta)+\frac{lR_0\omega_{co}}{L^2 u\sqrt{\kappa}\vartheta_\parallel}C(\theta)\right]\frac{\cos\theta\sqrt{1+3\sin^2\theta}}{\sqrt{1-\mu b(\theta)}}\,d\theta, \tag{28}$$

$$\hat{D}_{p,l}^n(u,\mu) = \int_{-\theta_l}^{\theta_l}\cos\left[\frac{n\pi}{\lambda_o}\lambda(\theta)-p\frac{2\pi}{\tau_b}\tau(\theta)-\frac{lR_0\omega_{co}}{L^2 u\sqrt{\kappa}\vartheta_\parallel}C(\theta)\right]b(\theta)\frac{\cos\theta\sqrt{1+3\sin^2\theta}}{\sqrt{1-\mu b(\theta)}}\,d\theta, \tag{29}$$

$$u = \frac{v}{\sqrt{\kappa}\,\vartheta_{\parallel}}, \qquad \hat{Z}_l(\mu) = \frac{R_0 L \tau_b}{2\pi\sqrt{\kappa}\,\vartheta_{\parallel}}\left(\omega + l\,\frac{\omega_{co}}{L^3}\,\bar{b}\right). \tag{30}$$

It should be noted that, in the variables $(v, \mu)$, there are identities for the phase coefficients: $G_{p,l}^n(v,\mu) = \hat{G}_{p,l}^n(v,\mu)$ and $D_{p,l}^n(v,\mu) = \hat{D}_{p,l}^n(v,\mu)$. Accordingly, the wave-particle resonance conditions are also independent of the steady-state distribution functions of the trapped particles and can be written as

$$\omega + l\,\frac{\omega_{co}}{L^3}\,\bar{b}(\mu) = p\,\frac{2\pi\,v}{R_0 L \tau_b(\mu)} \tag{31}$$

involving the wave frequency $\omega$, the bounce averaged cyclotron frequency $\omega_{co}\bar{b}(\mu)/L^3$ and the bounce frequency of the trapped particles $2\pi v/[R_0 L \tau_b(\mu)]$ with the given energy $v$ and pitch angle $\mu$ at the given (by $L$) magnetic field line, where $l = 0,\pm 1,...$ and $p = 0,\pm 1,...$ are the numbers of the cyclotron and bounce resonances, respectively.

In the case of the bi-Lorentzian distribution functions, the dispersion equation for the cyclotron waves will be the same as Eq. (22), where the new transverse dielectric tensor components, Eq. (27), should be included. For specified values of the ambient magnetic field line $L$, and the parameters $(\kappa, T_{\parallel\sigma}, T_{\perp\sigma}, N_\sigma, M_\sigma, e_\sigma)$ that describe each particle species, an iterative scheme can me used to solve Eq. (22) for complex wave frequency for a prescribed real values of the wave number $k_{\parallel} = n\pi/\lambda_o$.

## Conclusion

In this paper, we have derived the dispersion equations (22) for field-aligned cyclotron waves in an axisymmetric dipole magnetospheric plasmas with both the bi-Maxwellian (6) and bi-Lorentzian (24) distribution functions. To evaluate the contribution of the trapped particles to the transverse current density components the Vlasov equation is solved using a standard method of switching to new variables (2) associated with conservation integrals of particle energy, magnetic moment and equation of the geomagnetic field line; the new time-like variable $\tau(\theta)$, Eq.(12), is introduced (instead of the geomagnetic latitude angle $\theta$) to describe the bounce-periodic motion of the trapped particles along the geomagnetic field; the perturbed electric field and current density components are Fourier-decomposed over the length of the geomagnetic field lines, Eqs. (15). As a result, the transverse permittivity elements, Eqs. (18, 27) are expressed by summation of the bounce-resonant terms including the double integration in velocity space, the resonant denominators, and the corresponding phase coefficients. Due to magnetic field nonuniformity, the bounce-resonance conditions for trapped particles in magnetospheric plasmas are entirely different from ones in the straight magnetic field; the whole spectrum of the electric field is present in the given current density harmonic; the left- and right-hand polarised waves are coupled. To have some analogy with the linear theory of cyclotron waves in the straight magnetic field, there were assumed that the $n$-th harmonic of the electric field gives the main contribution to the $n$-th harmonic of the current density and the connection of the left- and right-hand waves is small. In this case, the dispersion equation for cyclotron waves has the simplest form and is suitable to analyse the instabilities of both the electron- and ion-cyclotron waves accounting for the cyclotron and bounce resonances.

*Acknowledgements:* This research was supported by CNPq of Brazil (Conselho Nacional de Desenvolvimento Científico e Tecnológico, project PCI-LNCC/MCT 382042/04-2).